\documentclass[conference]{IEEEtran}

\IEEEoverridecommandlockouts    

\usepackage{cite}
\usepackage{amsmath,amssymb,amsfonts}
\usepackage{graphicx}
\usepackage{textcomp}
\usepackage{acro}
\usepackage{url}
\usepackage{multirow}
\usepackage{stfloats}
\usepackage[caption=false, font=footnotesize]{subfig}
\usepackage{algorithm}
\usepackage{algpseudocode}
\usepackage{xcolor}
\usepackage{hyperref}
\usepackage{bbold}
\usepackage{titlesec}

\usepackage{pgf} 
\usepackage{booktabs}

\title{\vspace{0.1in}\LARGE \bf Scalable Vehicle Charging Coordination for Online Operations:\\A Column Generation Approach}

\author{
	\parbox{\textwidth}{\centering
	Joas Kahlert$^1$, Ruiting Wang$^1$, Gabrijela Obradović$^2$, Henrik Svärd$^2$, Jonas Mårtensson$^1$}
	\thanks{$^{1}$Decision and Control Systems, KTH Royal Institute of Technology, Stockholm, Sweden, {\tt\footnotesize kahlert@kth.se, ruiting@kth.se, jonas1@kth.se.}}%
    \thanks{$^{2}$Scania CV, TRATON GROUP, Södertälje, Sweden, {\tt\footnotesize gabrijela. obradovic@scania.com, henrik.svard@scania.com.}}
}

\DeclareAcronym{BET}{
  short = BET ,
  long  = Battery Electric Truck,
  short-plural = s 
}

\DeclareAcronym{HoS}{
  short = HoS ,
  long  = Hours of Service,
  short-plural = s 
}

\DeclareAcronym{SoC}{
  short = SoC ,
  long  = State of Charge,
  short-plural = s 
}

\DeclareAcronym{MIP}{
  short = MIP ,
  long  = Mixed Integer Program,
  short-plural = s 
}

\DeclareAcronym{MILP}{
  short = MILP ,
  long  = Mixed Integer Linear Program,
  short-plural = s 
}

\DeclareAcronym{VSP}{
  short = VSP,
  long  = Vehicle Scheduling Problem,
  short-plural = s 
}

\DeclareAcronym{VRP}{
  short = VRP,
  long  = Vehicle Routing Problem,
  short-plural = s 
}

\DeclareAcronym{PDF}{
  short = PDF ,
  long  = Probability Density Function,
  short-plural = s 
}

\DeclareAcronym{FCFS}{
  short = FCFS ,
  long  = First Come First Served,
  short-plural = s 
}

\DeclareAcronym{ODE}{
  short = ODE ,
  long  = Ordinary Differential Equation,
  short-plural = s 
}

\DeclareAcronym{EV}{
  short = EV ,
  long  = Electric Vehicle,
  short-plural = s 
}

\DeclareAcronym{MCS}{
  short = MCS,
  long  = Megawatt Charging System,
  short-plural = s 
}

\DeclareAcronym{CCS}{
  short = CCS,
  long  = Combined Charging System,
  short-plural = s 
}

\DeclareAcronym{MP}{
  short = MP,
  long  = Master Problem,
  short-plural = s 
}

\DeclareAcronym{RMP}{
  short = RMP,
  long  = Restricted Master Problem,
  short-plural = s 
}

\DeclareAcronym{SP}{
  short = SP,
  long  = Sub Problem,
  short-plural = s 
}

\DeclareAcronym{LC}{
  short = LC,
  long  = Label Correcting,
  short-plural =  
}

\DeclareAcronym{CG}{
  short = CG,
  long  = Column Generation,
  short-plural =  
}

\DeclareAcronym{ECV}{
  short = ECV,
  long  = Electric Commercial Vehicle,
  short-plural =  
}

\DeclareAcronym{EVCSP}{
  short = EVCSP,
  long  = Electric Vehicle Charging Scheduling Problem,
  short-plural =  
}
\algrenewcommand{\algorithmiccomment}[1]{\hfill\textcolor{gray}{// #1}} 

\titlespacing*{\section}{0pt}{0.8ex}{0.5ex} 
\titlespacing*{\subsection}{0pt}{0.6ex}{0.8ex}

\begin{document}

\maketitle

\thispagestyle{empty}
\pagestyle{empty}

\begin{abstract}
We consider an electric vehicle charging scheduling problem for long-haul freight transport. To mitigate unforeseen waiting times caused by congestion and limited charger capacity, we consider a coordinated scheduling approach that prioritizes scalable, computationally efficient scheduling for ongoing or online operations. We formulate the problem as a set partitioning model and develop a column generation framework with a label-correcting algorithm to solve it efficiently. We assess the scalability of our algorithm through both theoretical complexity analysis and empirical measurements, highlighting the role of feasibility and dominance pruning in limiting label growth. Finally, using freight telematics data from Sweden, we show that the proposed method computes optimized schedules for realistic instances with 1000 vehicles within seconds and reduces vehicle operation downtime by 78.6\% compared to uncoordinated scheduling.
\end{abstract}


\section{Introduction}
The electrification of the long-haul transport sector has gained significant momentum in response to the growing demand for sustainable mobility solutions \cite{iea_2026}. This transition is driven not only by regulatory pressures, such as EU directives and national transport policies, but also by evolving customer preferences and rising public expectations \cite{iea_2026}.  

Compared with diesel-powered alternatives, \acp{EV} introduce several operational challenges, including shorter driving range, long charging times, and limited charging capacity at charging stations. Given these limitations, electric fleet operations become interdependent through shared but limited charging resources. Operating a fleet of \acp{EV} therefore requires more careful planning, with vehicle rosters and delivery schedules often assigned in an upstream planning stage, typically a day ahead of operations. In the literature, such problems are commonly studied as variants of the electric \ac{VRP} or \ac{VSP} \cite{desrochers_1992}, \cite{desaulniers_2016}, \cite{froger_2022}, \cite{lam_2022}, \cite{wang_2021} and \cite{eveborn_2026}.

Large-scale routing and scheduling problems are often computationally challenging \cite{desaulniers_2016}. A common strategy is to exploit favorable problem structure through decomposition. In this context, \ac{CG} provides a promising approach for solving the resulting formulations efficiently.

An early contribution in this field \cite{desrochers_1992} applied \ac{CG} to the \ac{VRP} with time windows, demonstrating how the problem can be formulated as a set partitioning problem and its linear programming relaxation solved to optimality using \ac{CG}. In line with the growing importance of \acp{EV} this formulation has been extended by studies such as \cite{desaulniers_2016} to also consider \ac{EV} range, battery, and recharge limitations. More recent contributions, \cite{froger_2022}, \cite{lam_2022} and \cite{klein_2023}, incorporate additional complexities such as nonlinear charging curves, battery degradation, and recognize the importance of limited charging station capacities. Other works, including \cite{wang_2021}, \cite{eveborn_2026} and \cite{parmentier_2023}, improve the computational efficiency of \ac{CG} for \acp{VSP} through more sophisticated \ac{CG} techniques.

Much of the existing literature focuses on classical \ac{VRP} or \ac{VSP} formulations for day-ahead planning, which provide initial routes and schedules but often assume idealized operating conditions. In practice, however, unforeseen events and deviations from the original plan may require interdependent \ac{EV} schedules to be revised and re-coordinated at short notice.   

To address this need, we focus on generating up-to-date coordinated charging schedules for a fleet of \acp{EV} in an online setting, given ongoing or pre-assigned delivery schedules determined by an upstream planning problem. We refer to this problem as the \ac{EVCSP}. Routing decisions are assumed to be made independently by individual \acp{EV} and therefore fall outside the scope of the scheduling coordinator. Accordingly, we consider scheduling interactions along a single road segment or highway, referred to as a charging corridor, while noting that the proposed concept can be readily extended to broader road networks. Our aim is to develop a computationally efficient method capable of optimizing charging schedules within seconds. The method accounts for the current vehicle states and seeks to minimize fleet-wide charging overhead while satisfying operational and infrastructure constraints.

We previously investigated this problem using an alternative solution method in \cite{kahlert_2026}, which subsequently led to a collaboration with industrial experts presented in \cite{geijer_2026}. Both studies have served as important motivation for the present work. 

The contribution of this paper is summarized as follows: 
\begin{itemize}
    \item We formulate the fleet-wide \ac{VSP} as a set partitioning problem and solve it using an efficient \ac{CG} framework. Applied to a real-world case study in Sweden, the method generates high-quality charging schedules for up to $1000$ \acp{EV} in seconds.
    \item We derive the algorithm's theoretical time complexity and complement it with empirical scalability metrics, demonstrating favorable scaling in the considered case study. 
    \item We benchmark the \ac{CG} solution quality against an alternative \ac{MILP} formulation and use naive, uncoordinated scheduling to quantify the benefits of coordination.
\end{itemize}

We also make our codebase and computational studies 
publicly available at \href{https://github.com/joask9/evcsp.git}{https://github.com/joask9/evcsp.git}.
\label{sec:introduction}

\section{Problem Overview}

\subsection{Problem Description and Mathematical Notation}
\label{sec:problem}
In line with the work presented in \cite{kahlert_2026} and \cite{geijer_2026}, the \ac{EVCSP} for a fleet of \acp{EV} $n\in\mathcal N$, with pre-assigned delivery schedules that operate along a transport corridor with a sequence of installed charging stations $i\in\mathcal I$. Charging stations are assumed to be located close to the corridor and therefore only incur a minor charging overhead $\delta_i$. Stations are equipped with a limited number of chargers $C_{ih}$ of class $h\in\mathcal H$, where each charger class provides a charging power $P_h$. We use a common set of charger classes, $\mathcal H$, for all stations. If charger class $h$ is not available at station $i$, we instead set the capacity $C_{ih}=0$. Corridor characteristics are summarized as $g=(\{\delta_i\}_{i\in\mathcal I},\{C_{ih}\}_{(i,h)\in\mathcal I\times\mathcal H},\{P_h\}_{h\in\mathcal H})$.

\acp{EV} enter and exit the corridor at different points enumerated by an event index $\mathcal M_n=\{0,\ldots,M_n\}$. They consequently pass only an ordered subset of charging stations $i\in\mathcal I_n=\{i^n_0,\hdots,i^n_{M_n}\}\subseteq\mathcal I$ called a route. The \ac{EV} fleet is heterogeneous. Each \ac{EV} $n$ has a battery capacity $B_n$, a specified driving speed, and a power consumption. Based on these characteristics, the travel time and energy consumption from station $i_m^n\rightarrow i_{m+1}^n$ are deterministically computed as $\rho_{nm}$ and $\nu_{nm}$, respectively. To ensure \acp{EV} do not run out of battery, their \ac{SoC} $e_{m}$ must remain within the operational range $e_{m}\in[0,1]$, for all events $m\in\mathcal M_n$. \acp{EV} enter the corridor at time $t_n^\text{start}$ with initial \ac{SoC} $e_n^\text{start}$. In addition, \acp{EV} are also subject to EU \ac{HoS} rest-break regulations, which require the driver to take a rest break of $T^\text{rest}=45$ minutes in total after every $4.5$ hours of driving \cite{eu_road_transport_2026}. In our setting, the need for rests for \ac{EV} $n$ is prespecified based on the indicator $r_n\in\{0,1\}$. \ac{EV} characteristics are summarized as $v_n=(\mathcal I_n,B_n,\{\rho_{nm}\}_{m\in\mathcal M_n},\{\nu_{nm}\}_{m\in\mathcal M_n},t_n^\text{start},e_n^\text{start},r_n)$.

\subsection{Schedule Representation}
A charging schedule comprises a set of scheduling events that together form a time-slot table specifying which station and charger to wait, charge, and rest at. Timeslots are defined on a high-resolution time grid with step size $\Delta t$, where each time slot is indexed by $s\in\mathcal S=\{0,\hdots S\}$ and corresponds to the interval $[s\Delta t,(s+1)\Delta t]$. A single scheduling event $\Lambda_{m}=(h_{m},w_{m},\tau_{m})$ with event index $m$ specifies whether to visit $\chi_{m}\in\{0,1\}$, whether to rest $\theta_{m}\in\{0,1\}$, the assigned charger class $h_{m}\in\mathcal H$, the waiting time $w_{m}\in\mathcal W=\{0,\Delta t,\ldots,W\}$, and the charging time $\tau_{m}\in\mathcal T=\{\underline T,\underline T+\Delta t,\ldots, \overline T\}$.

To reduce the state space of the scheduling event, we have excluded the explicit visit and resting decisions $\chi_{m}$ and $\theta_{m}$ from $\Lambda_{m}$. Instead, we assume the station is visited when time spent at the station is non-zero, 
\begin{equation*}
    \label{eq:charge}
    \chi_{m}=\mathbb 1(w_{m}+\tau_{m}> 0).
\end{equation*}
When the total time spent at a station is greater than the required rest time, rest can occur
\begin{equation*}
    \label{eq:rest}
    \theta_{m}=\mathbb 1(w_{m}+\tau_{m}\geq T^\text{rest}).
\end{equation*} 
With this formulation, visiting and resting are explicitly determined, based only on the lengths of the waiting and charging intervals.

A complete charging schedule $p$ is obtained by assigning some scheduling event $\Lambda_{m}$ for each event index $m\in\mathcal M$ according to,
\begin{equation}
    \label{eq:schedule}
    p=\{\Lambda_{m}\}_{m\in\mathcal M}.
\end{equation}
The cost $c_{p}$ of $p$ is defined as the sum of schedule-related overhead $t_{m}$ at each event $m$,
\begin{equation}
    \label{eq:cost}
    c_{p}:=\sum_{m\in\mathcal M}t_{m},\quad t_{m}=\chi_{m}\delta_{i_m}+w_{m}+\tau_{m}.
\end{equation}

\label{sec:schedule}

\subsection{Problem Definition}
Let $\mathcal P_n$ be the set of feasible schedules for \ac{EV} $n$ and define $\lambda_{np}$ to be the binary variable indicating whether schedule $p\in\mathcal P_n$ is selected for \ac{EV} $n$. We consider an \ac{EV} schedule $p\in\mathcal P_n$ feasible if it satisfies: 1) the \ac{SoC} limits $e_{m}\in[0,1],\ \forall m\in\mathcal M_n$, and 2) the \ac{HoS} regulations $\sum_{m\in\mathcal M_n}\theta_{m}\geq r_n$. 

To ensure that a combination of schedules respects the charging-capacity limits $C_{ih},\ \forall (i,h,s)\in\mathcal I\times\mathcal H\times\mathcal S$, let $b^{ihs}_{p}$ be the occupancy indicator 
\begin{equation}
\label{eq:b}
    b_{p}^{ihs}=
    \begin{cases}
    1, & \begin{aligned}
        &\text{if charging schedule $p$ assigns charger}\\
        &\text{class $h$ at station $i$ at time $s$}
    \end{aligned}\\
    0, & \text{else}
    \end{cases}
\end{equation}

Given an \ac{EV} fleet $n\in\mathcal N$, we assign one feasible charging schedule $p\in\mathcal P_n$ per \ac{EV} that satisfies charging-capacity limits, minimizing total schedule cost. This can be modeled as a set partitioning problem over the sets of schedules $\mathcal P_n$.
\begin{subequations}
\label{eq:mp}
\begin{align}
    \!\min&\sum_{n\in\mathcal N}\sum_{p\in\mathcal P_n}\!c_{p}\lambda_{np}\label{eq:mp_obj}\\
    \text{s.t.}\ &\sum_{{p}\in\mathcal P_n}\!\lambda_{np}=1\qquad\qquad\quad[\mu_n]\quad \forall n\in\mathcal N\label{eq:one_column}\\
    &\sum_{n\in\mathcal N}\sum_{{p}\in\mathcal P_n}\!b^{ihs}_{p} \lambda_{np}\leq C_{ih}\ [\pi_{ihs}]\ \forall (i,h,s)\!\in\!\mathcal I\!\times\!\mathcal H\!\times\!\mathcal S\label{eq:mp_coupling}\raisetag{8pt}\\
    &\lambda_{np}\in\{0,1\}\qquad\qquad\qquad\qquad\forall(n,{p})\!\in\!\mathcal N\!\times\!\mathcal P_n\label{eq:bin_column}
\end{align}
\end{subequations}

The objective \eqref{eq:mp_obj} minimizes the total cost of the selected schedules. The set partitioning constraint \eqref{eq:one_column} ensures that exactly one schedule is selected for each \ac{EV}. Constraint \eqref{eq:mp_coupling} enforces the fleet-wide charging-capacity limits for all time slots. The left-hand side of the constraint gives the number of simultaneous charging sessions at each station and charger type, which cannot exceed the available capacity $C_{ih}$. $\mu_n\in\mathbb R$ and $\pi_{ihs}\leq0$ are dual variable associated with Constraints \eqref{eq:one_column} and \eqref{eq:mp_coupling}. \eqref{eq:bin_column} enforces binary schedule selection.
\label{sec:def}

\section{Column Generation}
In this work, we focus primarily on implementing \ac{CG} for the problem under study, rather than on the underlying theoretical framework. Readers seeking a more comprehensive introduction to \ac{CG} are referred to \cite{desrosiers_2026}.

\label{sec:CG}


\subsection{Column Generation Framework}
\ac{CG} is a decomposition technique for optimization problems with a large number of variables where a solution can be naturally represented as a combination of well-defined configurations, referred to as columns. A column may, for example, represent a pattern, route, or, in our case, a charging schedule $p$. Columns are decision variables in the top-level optimization problem, referred to as the \ac{MP} in \eqref{eq:mp}.

Typically, the number of possible columns is too large to solve the \ac{MP} directly. Therefore, \ac{CG} works with a reduced version of the \ac{MP} called the \ac{RMP}. The \ac{RMP} is obtained by replacing the set $\mathcal P_n$ in \eqref{eq:mp} with a restricted subset $\widetilde{\mathcal P}_n\subset\mathcal P_n$ and relaxing the integrality constraint \eqref{eq:bin_column} $\lambda_{np}\in[0,1]$. The resulting \ac{RMP} is a relatively small linear program and can easily be solved with commercial solvers such as Gurobi \cite{gurobi_2026}. 

Solving the \ac{RMP} provides the dual values $\mu_n$ and $\pi_{ihs}$, which are used to find new columns with negative reduced cost 
\begin{equation}
\label{eq:red_cost}
    \overline c_{p}=c_{p}-\mu_n-\sum_{i\in\mathcal I_n}\phi_{p}^i,\quad \phi^i_{p}=\sum_{h\in\mathcal H}\sum_{s\in\mathcal S}b^{ihs}_{p}\pi_{ihs},
\end{equation}
that might improve the \ac{RMP} objective \cite{desrosiers_2026}. Here $\phi^i_{p}$ is the charging penalty defined as the accumulated cost over all charging timeslots at station $i$. New columns $p\in\mathcal P_n$ are generated  by solving the \ac{SP}
\begin{equation}
\label{eq:sp}
    \hat{\overline c}_n=\min_{p\in\mathcal P_n}\overline c_{p},
\end{equation}
which returns the column $\hat p\in\mathcal P_n$ with the minimum reduced cost $\hat{\overline c}_n$ relative to the current \ac{RMP} solution. If $\hat{\overline c}_n<-\varepsilon$, for some small tolerance $\varepsilon>0$, the corresponding column $\hat p$ is added to the restricted column set $\widetilde{\mathcal P}_n$, and the \ac{RMP} is resolved with the updated column sets. This procedure is repeated until no improving column can be found. 

To obtain a final integer solution, we solve the \ac{MP} over the generated column sets $\widetilde{\mathcal P}_n$. In this way, \ac{CG} can often produce high-quality solutions efficiently while generating only a small fraction of all possible columns. The \ac{CG} algorithm is summarized in Algorithm \ref{alg:cg}.

\begin{algorithm}[t]
\caption{Column Generation Algorithm}
\label{alg:cg}
\begin{algorithmic}[1]
    \Require EV and corridor characteristics $v_n$ and $g$
    \Ensure Optimal columns $\hat{\mathcal P}$ and \ac{MP} objective $\hat{z}$
    \State  Initialize $\hat{\overline c}_n,\hat p\gets$ solve SP:  LC$(v_n,0,-\infty),\ \widetilde{\mathcal P}_n \gets \{\hat p\}, \forall n\in\mathcal N$
    \While{$\exists n\in\mathcal N:\ \hat{\overline c}_n<-\varepsilon$}\Comment{$\exists$ improving column?}
        \State $\mu_n,\pi_{ihs}\gets$ solve \ac{RMP} with $\widetilde{\mathcal P}_n$\Comment{def. \eqref{eq:mp}}
        \For{$n\in\mathcal N$}
            \State $\hat{\overline c}_n,\hat p\gets$ solve \ac{SP}:  LC$(v_n,\mu_n,\pi_{ihs})$\Comment{alg. \ref{alg:lc}}
            \If{$\hat{\overline c}_n < -\varepsilon$}\Comment{if improving column}
                \State $\widetilde{\mathcal P}_n \gets \widetilde{\mathcal P}_n \cup \{\hat p\}$\Comment{add column}
            \EndIf
        \EndFor
    \EndWhile
    \State $\hat z, \hat{\mathcal P}\gets$ Solve \ac{MP} with $\widetilde{\mathcal P}_n$   
\end{algorithmic}
\end{algorithm}

\label{sec:cg}

\subsection{Label Correcting}
To solve the \ac{SP} \eqref{eq:sp} for \ac{EV} $n$, we employ a \ac{LC} Algorithm \ref{alg:lc}, that takes advantage of the sequential structure of a column \eqref{eq:schedule} by decomposing it into a sequence of connected scheduling events $\Lambda_{m}$ over the event set $m\in\mathcal M_n$. Starting from $m=0$, \ac{LC} advances the partial schedule $p_m=\{\Lambda_{0},\hdots,\Lambda_{m-1}\}$ by repeatedly extending $p_{m+1}=p_m\cup\{\Lambda_{m}\}$ with some event $\Lambda_{m}$. After the last extension at $m=M_n$, we obtain a complete column $p_{M_n}=p\in\mathcal P_n$, provided that $p$ is feasible for the assigned \ac{EV}.

Each assigned partial schedule $p_m$ is associated with a multi-dimensional vector called a label, denoted as,
\begin{equation*}
    p_m:L=(\ell^\text{time},\ell^\text{SoC},\ell^\text{rest},\ell^\text{rc})\in\mathcal L_{m}.
\end{equation*}
A label tracks (time, \ac{SoC}, rest, reduced cost) as \ac{EV} resources at event $m$. Labels that have advanced to the same stage $m$ are defined to belong to the set $\mathcal L_{m}$.

Labels are updated by an extension process $L'=f_{nm}(L,\Lambda,v)$ \eqref{eq:extension}. The label extension function $f_{nm}$ maps a label $L\in\mathcal L_{m}$ to a new label $L'\in\mathcal L_{m+1}$ following the schedule extension $p_m\rightarrow p_{m+1}$ by updating label resources based on the current label state $L$, the specific charging event $\Lambda_m=(h_m,w_m,\tau_m)$, and \ac{EV} characteristics $v_n$. We define,
\begin{equation}
\label{eq:extension}
\begin{aligned}
    f_{nm}(L,\Lambda,v)=L&+(t_{m},\Delta e_{nm},\theta_{m},t_{m}-\phi_{p_m}^{i^n_m})\\
    &+(\rho_{nm},-\nu_{nm},0,0),
\end{aligned}
\end{equation}
where $t_{m}$ is the overhead from \ref{eq:cost} and $\phi_{p_m}^{i^n_m}$ is the charging penalty from \ref{eq:red_cost}. The first term distributes the charging overhead $t_{m}$, amount of \ac{SoC} charged $\Delta e_{nm}=\frac{\tau_{m}P_{h_{m}}}{B_n}$, rest $\theta_{m}$, and the objective contribution $t_m-\phi_{p_m}^{i^n_m}$ from charging event $\Lambda_{m}$ over the label resources. The second term is deterministic and updates resources due to traveling from $i^n_m\rightarrow i^n_{m+1}$. 

To ensure that a newly generated schedule extension $p_{m+1}$ remains feasible for \ac{EV} $n$, its label representation $p_{m+1}:L'$ must be verified to satisfy the \ac{EV} schedule feasibility constraints,
\begin{subequations}
\label{eq:feasibility}
\begin{align}
    &0\leq\ell^\text{SoC}\leq 1&\forall L\in\mathcal L_{m},\label{eq:feasibility:soci}\\ 
    &r_n\leq\ell^\text{rest}& \forall L\in\mathcal L_{M_n}. \label{eq:feasibility:rest}
\end{align}
\end{subequations}
We integrated these conditions directly into the \ac{LC} algorithm by allowing only extensions that guarantee a feasible resulting label. For this purpose, let 
\begin{equation}
\label{eq:omega}
    \Omega_{nm}(L)=\{(h,w,\tau)\vert h\in\mathcal H,w\in\mathcal W,\tau\in\mathcal T_{nmh}(L)\},
\end{equation}
be the set of all feasible scheduling events $\Lambda_{m}\in\Omega_{nm}(L)$ that will guarantee the extension $L'=f_{nm}(L,\Lambda,v)$ to satisfy \eqref{eq:feasibility:soci}. We recall that $\mathcal W=\{0,\Delta t,\ldots,W\}$, and $\mathcal T=\{\underline T,\underline T+\Delta t,\ldots, \overline T\}$. By setting the charging bounds according to
\begin{equation}
\label{eq:omegabounds}
    \overline T_{nmh}\!=\!\max(0,\nu_{nm}-\ell^\text{SoC})\frac{B_n}{P_h},\ \, \underline T_{nh}\!=\!(1-\ell^\text{SoC})\frac{B_n}{P_h},
\end{equation}
we ensure that the charging length $\tau_{m}\in\mathcal T_{nmh}(L)$ adds a feasible amount of energy $\Delta e_{nm}$ such that the extended \ac{SoC} $\ell'{}^\text{SoC}=\ell^\text{SoC}+\Delta e_{nm}-\nu_{nm}$ is within the specified bounds \eqref{eq:feasibility:soci}. Note that it if $\overline T_{nmh}<\underline T_{nh}$, no feasible extension exists, $\Omega_{nm}(L)=\emptyset$. The rest condition \eqref{eq:feasibility:rest} only needs to be checked after the last extension $m=M_n$. 

Starting from a root label with initial values $p_0:L^0=(t_n^\text{start},e_n^\text{start},0,-\mu_n)$ and repeatedly applying $f_{nm}(L,\Lambda,v), \ \forall\Lambda\in\Omega_{nm}(L),\ \forall m\in\mathcal M_n$, the extension $f_{nm}$ can in principle enumerate all possible columns. However, the number of columns may be very large in practice. To reduce the number of labels to explore, we apply a dominance rule to discard labels that are provably inferior to others. The dominance rule we employ is 
\begin{equation}
\label{eq:dominance}
\begin{gathered}
    L_\alpha\succ L_\beta\quad\text{if}\quad
    L_\alpha,L_\beta\in\mathcal L_{m}\ \wedge\\
    \begin{aligned}
        &\!\left[
        (\ell_\alpha^{\text{time}}\!\leq\!\ell_\beta^{\text{time}})\!\wedge\!
        (\ell_\alpha^{\text{SoC}}\!\geq\!\ell_\beta^{\text{SoC}})\!\wedge\!
        (\ell_\alpha^{\text{rest}}\!\geq\!\ell_\beta^{\text{rest}})\!\wedge\!
        (\ell_\alpha^{\text{rc}}\!\leq\!\ell_\beta^{\text{rc}})
        \right]\wedge\\
        &\!\left[
        (\ell_\alpha^{\text{time}}\!<\!\ell_\beta^{\text{time}})\!\vee\!
        (\ell_\alpha^{\text{SoC}}\!>\!\ell_\beta^{\text{SoC}})\!\vee\!
        (\ell_\alpha^{\text{rest}}\!>\!\ell_\beta^{\text{rest}})\!\vee\!
        (\ell_\alpha^{\text{rc}}\!<\!\ell_\beta^{\text{rc}})
        \right].
    \end{aligned}
\end{gathered}
\end{equation}

This condition compares two labels, $L_\alpha,L_\beta\in\mathcal L_{m}$. In terms of the relevant resources, clearly, earlier arrival, higher \ac{SoC}, completed rest, and lower reduced cost are preferable characteristics for any partial schedule. Hence, we say that label $L_\alpha$ dominates ($\succ$) label $L_\beta$ if $L_\alpha$ is at least as good as $L_\beta$ in all resources and strictly better in at least one. 

\begin{algorithm*}[t]
\caption{ Label Correcting Algorithm}
\label{alg:lc}
\begin{algorithmic}[1]
    \Require \ac{EV} characteristics $v_n$ and dual values $\mu_n,\pi_{ihs}$
    \Ensure Minimal cost $\hat{\overline c}_n$ and corresponding column $\hat p\in\mathcal P_n$ with occupancy indicator $b_{\hat p}^{ihs}$
    \State Initialize $L^0\gets(t_n^\text{start},e_n^\text{start},0,-\mu_n),\ \mathcal Q_0\gets\{L^0\},\ \mathcal Q_m\gets\emptyset\ \forall m\in\mathcal M_n\setminus\{0\},\ \mathcal L_m\gets\emptyset\ \forall m\in\mathcal M_n,\ \hat{\overline c}_n\gets\infty,\ \hat p\gets\varnothing $
    \While{$\bigcup_{m\in\mathcal M_n}\mathcal Q_m\neq\emptyset$}\Comment{have all labels been processed?}
        \State $m,L\gets$ pick unprocessed label from $\mathcal Q_m$
        \For{$\Lambda_{m}\in\Omega_{nm}(L)$}\Comment{ensure feasible extension, def. \eqref{eq:omega}, \eqref{eq:omegabounds}}
            \State $p_{m+1}\gets p_m\cup\{\Lambda_{m}\},\ L'\gets f_{nm}(L,\Lambda,v)$\Comment{extend schedule and its label representation, def. \eqref{eq:extension}}
            \If{$\nexists L^*:L^*\!\succ\!L',\forall L^*\!\in\!\mathcal Q_{m+1}\cup\mathcal L_{m+1}$}\Comment{is the new label non-dominated? def. \eqref{eq:dominance}}
                \State $\mathcal D\gets \{L^*\vert L^*\prec L',\forall L^*\in\mathcal Q_{m+1}\}$\Comment{are any existing labels dominated? def. \eqref{eq:dominance}}
                \State $\mathcal Q_{m+1}\gets(\mathcal Q_{m+1}\setminus\mathcal D)\cup\{L'\}$\Comment{remove dominated labels $\mathcal D$ from $\mathcal Q_{m+1}$ and add the new one}
            \EndIf
        \EndFor
        \State $\mathcal L_m\gets\mathcal L_m\cup\{L\},\ \mathcal Q_m\gets\mathcal Q_m\setminus\{L\}$\Comment{add processed label to $\mathcal L_m$ and remove it from $\mathcal Q_m$}
    \EndWhile
    \State $\mathcal L_{M_n}\gets\{L\vert \ell^\text{rest}\geq r_n,L\in\mathcal L_{M_n}\}$\Comment{rest fulfilled? def. \eqref{eq:feasibility:rest}}
    \If{$\mathcal L_{M_n}\neq\emptyset$}\Comment{have any columns been found?}
        \State $\hat p:\hat L\gets\arg\min_{L\in\mathcal L_{M_n}} \ell^\text{rc}$\Comment{find column with minimal reduced cost among all representations $\mathcal L_{M_n}$, def. \eqref{eq:sp}}
        \State $\hat{\overline c}_n\gets\hat\ell^\text{rc}$\Comment{retrieve minimal reduced cost}
        \State $b_{\hat p}^{ihs}\gets$ retrieve occupancy indicator for $\hat p$\Comment{def. \eqref{eq:b}}
    \EndIf
\end{algorithmic}
\end{algorithm*}

\setlength{\textfloatsep}{8pt}

\label{sec:lc}

\section{Results and Discussion}

\subsection{Theoretical Scalability}
The time complexity of the \ac{CG} Algorithm \ref{alg:cg} is,
\begin{equation}
\label{eq:OCG}
    \mathcal O(\text{CG})=\mathcal O\left(\sum_{r\in\mathcal R}\left[\mathcal O(\text{RMP}_r)+\sum_{n\in\mathcal N}\mathcal O(\text{LC}_{rn})\right]\right).
\end{equation}
$r\in\mathcal R$ is the iteration index of the \ac{CG} algorithm and is assumed to be bounded by some constant $R=\vert\mathcal R\vert$, where $\vert\cdot\vert$ to denotes the cardinality of the set. For convenience of notation, we also define $I=\vert\mathcal I\vert$, $N=\vert\mathcal N\vert$ and $ H=\vert\mathcal H\vert$. In the following analysis, we assume that solving the \ac{RMP} can be done efficiently such that $\mathcal O(\text{RMP}_r)\ll\sum_{n\in\mathcal N}\mathcal O(\text{LC}_{rn})$ and therefore focus our attention on $\mathcal O(\text{LC}_{rn})$. 

To do so, we first determine the label propagation and growth dynamics based on the extension, feasibility, and dominance rules we established earlier. Now, consider the superset of label-independent scheduling events $\widehat\Omega_{nm}\supseteq\Omega_{nm}(L)$ as 
\begin{equation}
    \widehat\Omega_{nm}:=\Omega_{nm}(L=(\ell^\text{time},0,\ell^\text{rest},\ell^\text{rc})).
\end{equation} 
of size 
\begin{equation}
    \label{eq:sizeOmega}
    \vert\widehat\Omega_{nm}\vert=\vert\mathcal W\vert\sum_{h\in\mathcal H}\vert\mathcal T_{nmh}\vert=\frac{W}{\Delta t}\sum_{h\in\mathcal H}\frac{(1-\nu_{nm})B_n}{\Delta tP_h}.
\end{equation}
With this definition, we obtain the computable bound $\vert\widehat{\Omega}_{nm}\vert \geq \vert\Omega_{nm}(L)\vert$ on the number of new labels that can be generated from any label $L$ at each extension step. Following the procedure from Algorithm \ref{alg:lc} a candidate label set $\mathcal L_{nm}^r$ for \ac{EV} $n$, is extended over all scheduling events in $\widehat\Omega_{nm}$ forming the new set,
\begin{equation}
\label{eq:L+barbar}
    \overline{\overline{\mathcal L}}{}^r_{nm+1}:=\{L'\vert L'=f_{nm}(L,\Lambda,v),\forall L\in\mathcal L_{nm}^r,\forall\Lambda\in\widehat\Omega_{nm}\}.
\end{equation}
Infeasible labels are pruned from $\overline{\overline{\mathcal L}}{}^r_{nm+1}$ following the feasibility condition \eqref{eq:feasibility},
\begin{equation}
\label{eq:L+bar}
    \overline{\mathcal L}{}^r_{nm+1}:=\{L\vert L\text{ feasible},\forall L\in\overline{\overline{\mathcal L}}{}^r_{nm+1}\},
\end{equation}
and dominated labels are pruned from $\overline{\mathcal L}{}^r_{nm+1}$ by applying the pairwise dominance condition \eqref{eq:dominance}, 
\begin{equation}
\label{eq:L+}
    \mathcal L_{nm+1}^r:=\{L\vert L\nprec L^*,\forall L,L^*\in\overline{\mathcal L}{}^r_{nm+1}\}.
\end{equation}

The three step set transformation $\mathcal L_{nm}^r\rightarrow\overline{\overline{\mathcal L}}{}^r_{nm+1}\rightarrow\overline{\mathcal L}{}^r_{nm+1}\rightarrow\mathcal L_{nm+1}^r$ thus extends the candidate set $\mathcal L_{nm}^r$ to a new feasible and non-dominated candidate set $\mathcal L_{nm+1}^r$ for which the procedure can be repeated. 

Note that, in practice, our algorithm performs the step $\mathcal L_{nm}^r \rightarrow \overline{\mathcal L}{}^r_{nm+1}$ directly by extending each label only over the label-dependent event set $\Omega_{nm}(L)$ in \eqref{eq:omega}. Since these extensions are feasible by construction, no subsequent feasibility pruning is required. For simplicity of analysis, however, we continue to use the label-independent set $\widehat\Omega_{nm}$, for which feasibility pruning must be considered explicitly.

To measure the feasibility and dominance pruning effect, we define the pruning factors $\alpha_{nm}^r,\beta_{nm}^r\in[0,1]$ as
\begin{equation}
\label{eq:ab}
    \alpha_{nm}^r=\frac{\vert\overline{\mathcal L}{}_{nm+1}^r\vert}{\vert\overline{\overline{\mathcal L}}{}^r_{nm+1}\vert},\quad\beta_{nm}^r=\frac{\vert\mathcal L_{nm+1}^r\vert}{\vert\overline{\mathcal L}{}^r_{nm+1}\vert}. 
\end{equation} 

The extension step $\mathcal L_{nm}^r\rightarrow\overline{\overline{\mathcal L}}{}^r_{nm+1}$ in \eqref{eq:L+barbar}, will generate $\vert\widehat\Omega_{nm}\vert$ new labels per label $L\in\mathcal L_{nm}^r$. Since different label-event pairs may generate identical labels, the number of distinct labels after extension will satisfy the bound $\vert\overline{\overline{\mathcal L}}{}^r_{nm+1}\vert\leq\vert\mathcal L^r_{nm}\vert\vert\widehat\Omega_{nm}\vert$. Using the definition \eqref{eq:ab}, we obtain $\vert\mathcal L_{nm+1}\vert=\alpha_{nm}\beta_{nm}\vert\overline{\overline{\mathcal L}}{}^r_{nm+1}\vert$. If we define the effective propagation factor $\gamma_{nm}^r$ as the average number of new labels created per existing label at step $m$ of the extension procedure:
\begin{equation}
    \label{eq:gamma}
    \gamma_{nm}^r:=\alpha_{nm}^r\beta_{nm}^r\vert\widehat\Omega_{nm}\vert,
\end{equation}
then the recursive set extension has the following relation:
\begin{equation}
    \label{eq:recLextend}
    \begin{aligned}
    \vert\mathcal L_{nm+1}^r\vert
    &\leq\alpha_{nm}^r\beta_{nm}^r\vert\overline{\overline{\mathcal L}}{}^r_{nm+1}\vert\\
    &=\alpha_{nm}^r\beta_{nm}^r\vert\mathcal L_{nm}^r\vert\vert\widehat\Omega_{nm}\vert=\gamma_{nm}^r\vert\mathcal L_{nm}^r\vert.
    \end{aligned}
\end{equation}
Starting from the root label $\vert\{L^0\}\vert=\vert\mathcal L_{n0}^r\vert=1$ and unrolling the recursion gives 
\begin{equation}
    \label{eq:gammaprod}
    \vert\mathcal L_{nm}^r\vert=\prod_{k=0}^{m-1}\gamma_{nk}^r,
\end{equation}
labels retained at step $m$ of the extension procedure. Thus, avoiding exponential label growth requires this product to grow subexponentially with the extension index.

The computational effort for one set extension, $\mathcal L_{nm}^r\rightarrow\mathcal L_{nm+1}^r$ involves a label extension step followed by a two-step pruning procedure as established before. In the worst case, this requires 
\begin{equation}
    \underbrace{\vert\hat\Omega_{nm}\vert\vert\mathcal L_{nm}^r\vert}_{\mathcal O(1)\text{ label extensions}}\!\!+\!\!\underbrace{\vert\overline{\overline{\mathcal L}}{}^r_{nm+1}\vert}_{\mathcal O(1)\text{ feasibility checks}}\!\!+\!\!\underbrace{2\vert\overline{\mathcal L}{}^r_{nm+1}\vert^2}_{\mathcal O(1)\text{ dominance checks}}
\end{equation}

Completing the set extension $\mathcal L_{n0}^r\rightarrow\mathcal L_{nM_n}^r$ over all events $m\in\mathcal M_n$ we get
\begin{subequations}
 \label{eq:OLCrn}
    \begin{align}
        &\mathcal O(\text{LC}_{rn})\\
        &\!=\!\mathcal O\!\left(\sum_{m\in\mathcal M_n}\!\!\!\left[\vert\hat\Omega_{nm}\vert\vert\mathcal L_{nm}^r\vert\!+\!\vert\overline{\overline{\mathcal L}}{}^r_{nm+1}\vert\!+\!2\vert\overline{\mathcal L}{}^r_{nm+1}\vert^2\right]\!\right)\!\label{eq:OLCrn:a}\\
        &\!=\!\mathcal O\!\left(2\!\!\sum_{m\in\mathcal M_n}\!\!\!\left[\vert\hat\Omega_{nm}\vert\vert\mathcal L_{nm}^r\vert\!+\!(\alpha_{nm}^r\vert\widehat\Omega_{nm}\vert\vert\mathcal L_{nm}^r\vert)^2\right]\!\right)\!\label{eq:OLCrn:b}.
    \end{align}   
\end{subequations}
In step \eqref{eq:OLCrn:b}, we use the definition in \eqref{eq:ab} and the recurrence relation \eqref{eq:recLextend}.

Given the worst case $\mathcal M_n=\mathcal I$, $B=\max_{n\in\mathcal N}B_n$, $P=\min_{h\in\mathcal H}P_h$, $\nu=\min_{(n,m)\in\mathcal N\times\mathcal M_n}\nu_{nm}$, without feasibility and dominance pruning $\alpha^r_{nm}=\beta^r_{nm}=1$, we get the upper bound $\mathcal O(\text{LC}_{rn})\leq\mathcal O(\text{LC})$ with
\begin{subequations}
\label{eq:OLC}
    \begin{align}
        \mathcal O(\text{LC})&=\mathcal O\left(2\sum_{i\in\mathcal I}\left[\vert\hat\Omega\vert\vert\mathcal L\vert+(\vert\widehat\Omega\vert\vert\mathcal L\vert)^2\right]\right)\label{eq:OLC:a}\\
        &=\mathcal O\left(\sum_{i\in\mathcal I}\vert\hat\Omega\vert^{2i}\right)\label{eq:OLC:b}\\
        &\!\!\!\!\overset{\vert\hat\Omega\vert>1}{=}\!\!\mathcal O\left(\vert\hat\Omega\vert^{2I}\right)\label{eq:OLC:d}\\
        &=\mathcal O\left(\left(\frac{(1-\nu)HWB}{\Delta t^2 P}\right)^{2I}\right)\label{eq:OLC:e}.
    \end{align}
\end{subequations}
In step \eqref{eq:OLC:b}, we use the definitions \eqref{eq:gamma} and \eqref{eq:gammaprod}. The subsequent step follows from the geometric series formula, while step \eqref{eq:OLC:e} uses the definition \eqref{eq:sizeOmega}. Finally, we substitute the expression \eqref{eq:OLC:e} back into Equation \eqref{eq:OCG} to obtain,
\begin{equation}
    \mathcal O(\text{CG})\leq\mathcal O\left(RN\left(\frac{(1-\nu)HWB}{\Delta t^2 P}\right)^{2I}\right).\label{eq:OCGx}
\end{equation}
The final expression \eqref{eq:OCGx} describes the worst-case time complexity of Algorithm \ref{alg:cg}, disregarding feasibility and dominance pruning. This bound describes the general scaling behavior of the algorithm with respect to the problem parameters, rather than providing a tight characterization of its complexity. 
Specifically, the bound scales linearly with $R$ and $N$, as $\Delta t^{-4I}$ with the time-grid resolution, and as $P^{-2I}$ with charging power. The parameters $H$, $W$, $B$, and $1-\nu$ each scale to the power $2I$.

\label{sec:theoretical}

\subsection{Case Study}
\begin{figure*}[htbp]
\vspace{-6pt}
    \centering
    {\includegraphics[width=0.9\textwidth]{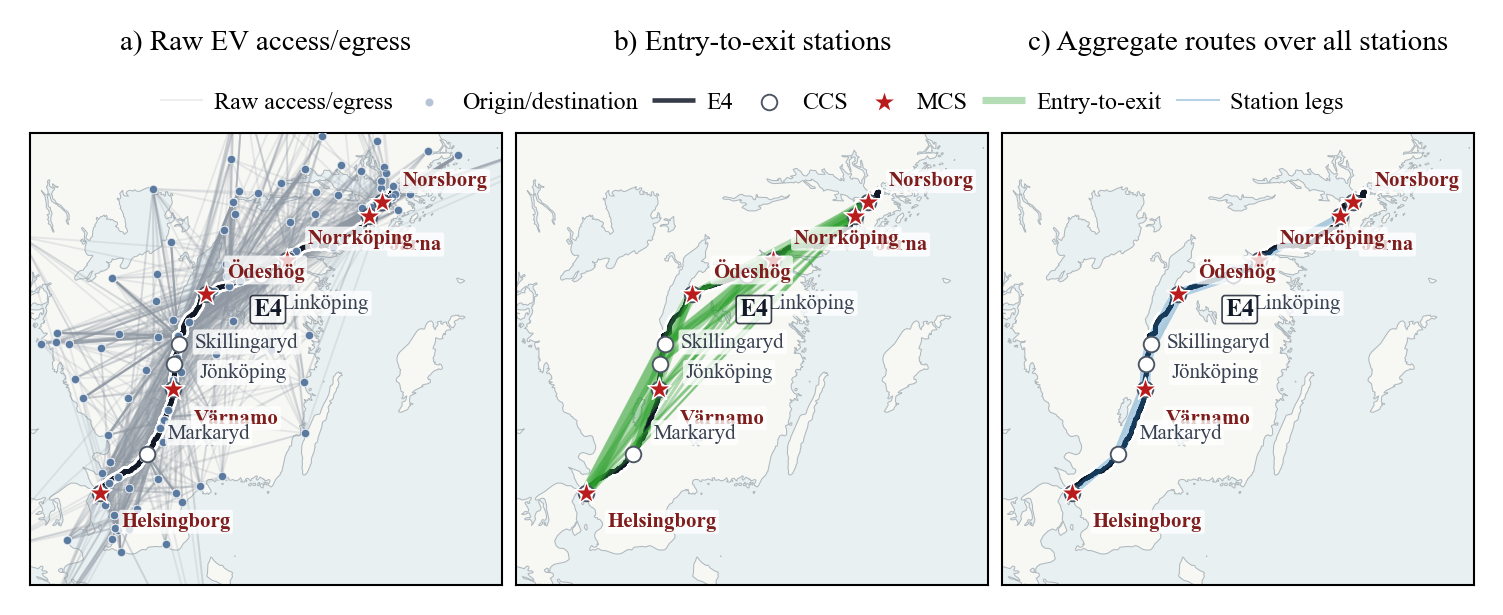}}
    \caption{Fleet-data transformation process from left to right. The panels illustrate the steps of a map-matching procedure in which raw \ac{EV} trip data is mapped onto a set of charging stations along the E4 in Sweden. a) Raw data; b) Entry and exit points $(i^n_0,i^n_{M_n})$; and c) Aggregated \ac{EV} routes $\mathcal I_n$ represented by all passed charging stations.}
    \label{fig:tripdata}
\vspace{-6pt}
\end{figure*}

To see what performance we can reasonably expect in a more realistic scenario, we apply the algorithm to a freight telematics dataset of a truck fleet. The dataset contains trajectory data for approximately $1200$ freight trucks observed over a one-day period, where each truck travels along the E4 highway in Sweden for a substantial part of its route. The dataset has been processed to extract \ac{EV} routes $\mathcal I_n$ and entry times $t_n^\text{start}$. See Figure \ref{fig:tripdata}. The rest indicator $r_n$ is pre-specified based on the dataset, depending on the total corridor travel time. However, because the dataset consists largely of diesel-powered trucks, we assume electrification parameters $B_n$, $e^\text{start}_n$, $\nu_{nm}$ to generate a scenario in line with today's standards \cite{scania_electric_trucks_2026}. 

In addition, we use a public charging map for trucks from \cite{skogforsk_2025} to localize currently available charging stations $\mathcal I$ within $2$ km of the E4 and determine travel time $\rho_{nm}$ and charging overhead $\delta_i$. We characterize charging options into two classes, $\mathcal H$, based on the power they provide: \ac{CCS} and \ac{MCS}. Charger type counts $C_{ih}$ per station are summarized in Table \ref{tab:station_data}. 

\begin{table*}[htbp]
\centering
\caption{E4 Charging Station Data ($C_{ih}$).}
\label{tab:station_data}
\footnotesize
\begin{tabular}{c|cccccccccc}
\hline
Stations & Norsborg & Järna & Norrköping & Linköping & Ödeshög & Jönköping & Skillingaryd & Värnamo & Markaryd & Helsingborg \\
\hline
$\#$ CCS    & 15& 6 & 14& 10& 15& 12& 8 & 5 & 6 & 15 \\
$\#$ MCS    & 8 & 1 & 1 & 0 & 1 & 0 & 0 & 1 & 0 & 1 \\
\hline
\end{tabular}
\vspace{-6pt}
\end{table*}

\label{sec:case}

\subsection{Analytical Scalability}
We complement the theoretical analysis with empirical results by applying the \ac{CG} algorithm to the \ac{EVCSP} with data from the case study. Using the full fleet of 1200 \acp{EV}, we compute coordinated schedules and record internal algorithm metrics during execution. The results are presented in the following Figure \ref{fig:L_gamma}.
\begin{figure}[htbp]
\vspace{-6pt}
    \centering
    \hspace*{-0.2cm}
    \includegraphics[width=1.00\linewidth]{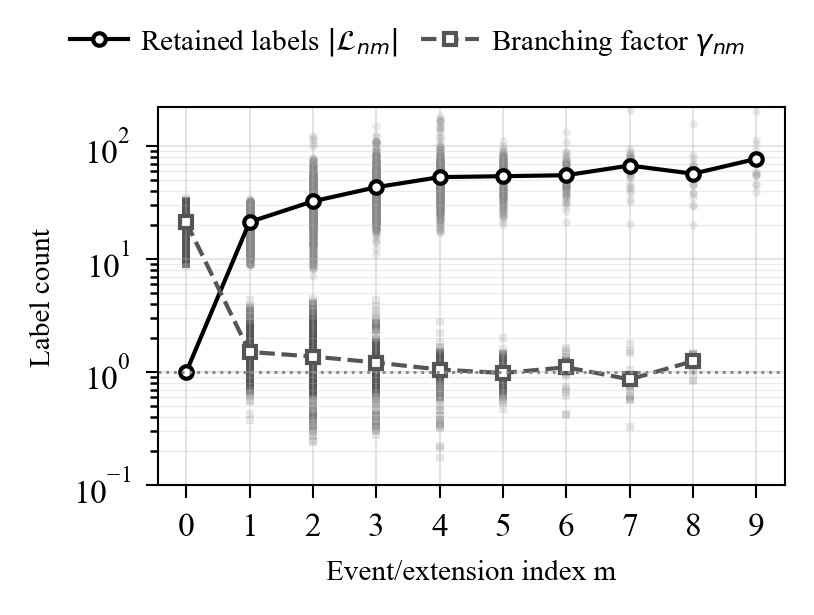}
    \caption{Propagation factor $\gamma_{nm}$ and the number of retained labels $|\mathcal L_{nm}|$ per extension step $m\in\mathcal M_n$. Each point represents one \ac{EV} $n$. The black lines indicate the average values across all \acp{EV} $n\in\mathcal N$ at each extension step.}
    \label{fig:L_gamma}
\vspace{-6pt}
\end{figure}

The figure shows the propagation factor $\gamma_{nm}$ and the number of retained labels $|\mathcal L_{nm}|$ per extension step $m$ averaged over all iterations $r\in\mathcal R$. Results show that the propagation factor is initially relatively large ($\gamma_{nm}>1$) during the early label extensions, resulting in the generation of many new labels. However, it decreases sharply and subsequently remains around $\gamma_{nm}\approx 1$. For some extensions, $\gamma_{nm}<1$, indicating that fewer than one label, on average, is retained per existing label. These observations suggest that the feasibility and dominance rules effectively limit label proliferation, maintaining a relatively stable number of propagated labels in the considered instance.

We next evaluate the computational efficiency of the proposed \ac{CG} approach on the truck-fleet dataset, comparing runtime and solution quality against two benchmark methods. A detailed description of the complete computational study is \href{https://github.com/joask9/evcsp.git}{publicly available}.

To evaluate solution quality, we compare against a continuous-time \textit{\ac{MILP}} formulation of the same \ac{EVCSP}, introduced in our previous work \cite{kahlert_2026}. The full formulation is provided in \cite{kahlert_2026}, with minor modifications to align it with the present framework.

Current electric trucks often operate independently, which may be sufficient during the early stages of freight electrification. However, as fleet sizes grow, this \textit{naive} approach may become inadequate. To represent such a scenario, we solve an uncoordinated \ac{EVCSP} by solving the \ac{SP} independently for each \ac{EV}, without enforcing the \ac{MP} linking constraints. The resulting schedules are individually optimal but ignore shared charging resources. We evaluate the approach through a chronological simulation in which vehicles follow their assigned schedules and wait in a first-come, first-served queue whenever all chargers are occupied.

\begin{figure}[htbp]
\vspace{-6pt}
    \centering
    \hspace*{-0.2cm}
    \includegraphics[width=1.00\linewidth]{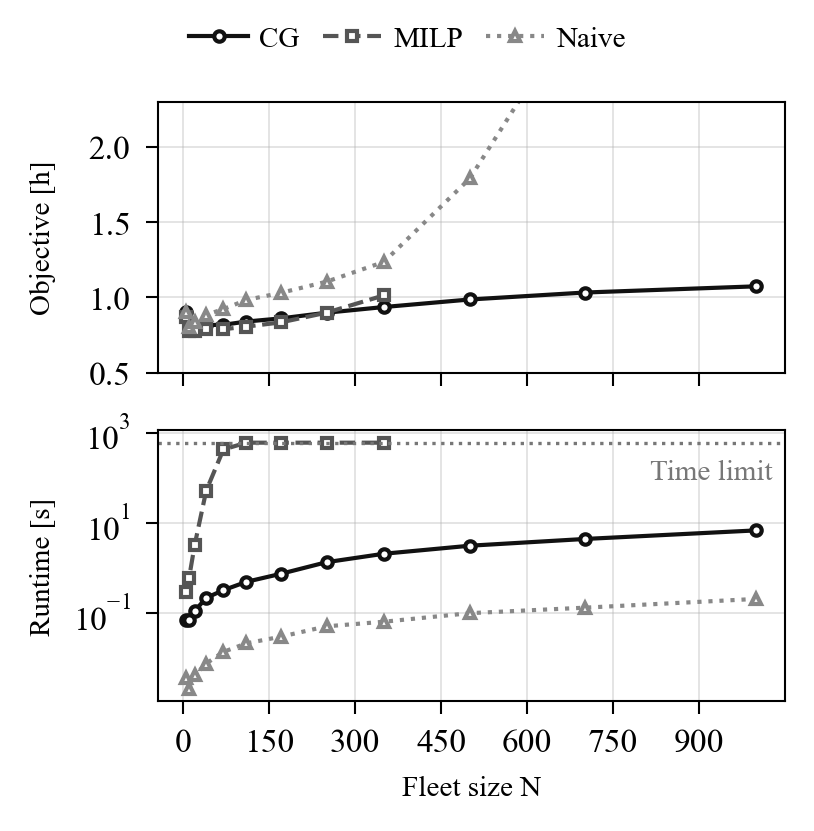}
    \caption{Benchmark results. Solution quality in terms of achieved objective (downtime per \ac{EV} $\frac{\hat z}{N}$) and computational runtime over varying fleet sizes $N$.}
    \label{fig:fig3}
\vspace{-6pt}
\end{figure}

We solve the \ac{EVCSP} on truck-fleet data and compare the \ac{MILP}, naive, and \ac{CG} approaches across increasing fleet sizes $\mathcal N$ to evaluate their performance under growing charging congestion. Experiments were performed on a Dell Latitude 7450 with an Intel Core Ultra 7 165U processor (12 cores, 14 threads) and 32 GB RAM, running 64-bit Windows 11. The implementation used Python 3.14.4 and Gurobi LLC v13.0.1 \cite{gurobi_2026} for all linear programs. The \ac{LC} algorithm was implemented in C++17 and compiled with GCC 15.2.0 using -O3. Reported runtimes are wall-clock times with a set time limit of $600$ seconds. We summarize the computational efficiency and solution quality in terms of the achieved objective in Table \ref{tab:benchmark_results} and Figure \ref{fig:fig3}. For interpretability, the objective is scaled by the total number of \acp{EV} $\frac{\hat z}{N}$. 

\begin{table}[htbp]
\centering
\caption{Benchmark results.}
\label{tab:benchmark_results}
\footnotesize
\begin{tabular}{r|cc| cc| cc}
\hline
\multirow{2}{*}{$N$}
& \multicolumn{2}{c}{CG}
& \multicolumn{2}{c}{MILP}
& \multicolumn{2}{c}{Naive} \\
\cline{2-7}
& Obj. & Time [s]
& Obj. & Time [s]
& Obj. & Time [s] \\
\hline
5    & 0.904 & 0.067 & 0.871 & 0.291   & 0.904 & 0.003 \\
10   & 0.804 & 0.069 & 0.776 & 0.578   & 0.806 & 0.002 \\
20   & 0.802 & 0.109 & 0.775 & 3.240   & 0.836 & 0.004 \\
40   & 0.816 & 0.213 & 0.792 & 51.249  & 0.884 & 0.007 \\
70   & 0.818 & 0.316 & 0.788 & 423.316 & 0.922 & 0.013 \\
110  & 0.839 & 0.484 & 0.806 & 600.000 & 0.983 & 0.021 \\
170  & 0.860 & 0.728 & 0.834 & 600.000 & 1.030 & 0.029 \\
250  & 0.898 & 1.324 & 0.898 & 600.000 & 1.104 & 0.050 \\
350  & 0.936 & 2.038 & 1.013 & 600.000 & 1.236 & 0.063 \\
500  & 0.986 & 3.075 & --    & --      & 1.791 & 0.097 \\
700  & 1.032 & 4.337 & --    & --      & 2.992 & 0.129 \\
1000 & 1.074 & 6.740 & --    & --      & 5.017 & 0.202 \\
\hline
\end{tabular}
\vspace{-6pt}
\end{table}

The naive approach is computationally efficient, but its solution quality deteriorates substantially for larger instances where schedules become increasingly interdependent. In such cases, we find that overhead time can increase by more than 5 times its initial value. The \ac{MILP} formulation avoids this problem by coordinated scheduling. However, it already fails to solve the $110$-trip instance to optimality within the specified time limit and is unable to find feasible solutions for instances larger than $350$ trips. The \ac{CG} method distinctively maintains both high solution quality and computational efficiency. For the 1000-trip instance, the \ac{CG} solution is obtained in 6.740 seconds and provides a 78.6\% reduction in downtime compared with uncoordinated or naive scheduling. 

\label{sec:analytical}

\section{Conclusion}  
This paper developed an effective \ac{CG} framework with a label-correcting algorithm for online charging-schedule optimization in large fleets of \acp{EV}. Theoretical analysis characterizes how computational effort scales with the problem parameters. Empirical results from the E4 case study demonstrate favorable scalability for practical problem instances. The method generates coordinated schedules for 1000 vehicles in seconds and reduces downtime by 78.6\% compared with uncoordinated scheduling, highlighting its potential for large-scale online scheduling applications.



\bibliographystyle{IEEEtran}
\bibliography{root} 
	
\end{document}